\documentclass[aip,reprint,superscriptaddress,longbibliography,floatfix]{revtex4-1}

\usepackage{amsmath,braket}
\usepackage{graphicx}
\usepackage{mathtools}
\usepackage{color}
\usepackage{xcolor}
\usepackage{setspace}
\usepackage{mathrsfs,amsmath}
\usepackage[pagewise]{lineno}
\makeatletter
\let\@fnsymbol\@fnsymbol@latex
\@booleanfalse\altaffilletter@sw
\makeatother

\begin{document}

\title{A bright and fast source of coherent single photons}

\author{Natasha Tomm}
\thanks{Natasha Tomm and Alisa Javadi contributed equally to this work}
\affiliation{Department of Physics, University of Basel, Klingelbergstrasse 82, CH-4056 Basel, Switzerland}

\author{Alisa Javadi}
\email{alisa.javadi@unibas.ch}
\affiliation{Department of Physics, University of Basel, Klingelbergstrasse 82, CH-4056 Basel, Switzerland}

\author{Nadia O.\ Antoniadis}
\affiliation{Department of Physics, University of Basel, Klingelbergstrasse 82, CH-4056 Basel, Switzerland}

\author{Daniel Najer}
\affiliation{Department of Physics, University of Basel, Klingelbergstrasse 82, CH-4056 Basel, Switzerland}

\author{Matthias C.\ L\"{o}bl}
\affiliation{Department of Physics, University of Basel, Klingelbergstrasse 82, CH-4056 Basel, Switzerland}

\author{Alexander R.\ Korsch}
\affiliation{Department of Physics, University of Basel, Klingelbergstrasse 82, CH-4056 Basel, Switzerland}
\affiliation{Lehrstuhl f\"{u}r Angewandte Festk\"{o}rperphysik, Ruhr-Universit\"{a}t Bochum, D-44780 Bochum, Germany}

\author{R\"{u}diger Schott}
\affiliation{Lehrstuhl f\"{u}r Angewandte Festk\"{o}rperphysik, Ruhr-Universit\"{a}t Bochum, D-44780 Bochum, Germany}

\author{Sascha R.\ Valentin}
\affiliation{Lehrstuhl f\"{u}r Angewandte Festk\"{o}rperphysik, Ruhr-Universit\"{a}t Bochum, D-44780 Bochum, Germany}

\author{Andreas D.\ Wieck}
\affiliation{Lehrstuhl f\"{u}r Angewandte Festk\"{o}rperphysik, Ruhr-Universit\"{a}t Bochum, D-44780 Bochum, Germany}

\author{Arne Ludwig}
\affiliation{Lehrstuhl f\"{u}r Angewandte Festk\"{o}rperphysik, Ruhr-Universit\"{a}t Bochum, D-44780 Bochum, Germany}

\author{Richard J.\ Warburton}
\affiliation{Department of Physics, University of Basel, Klingelbergstrasse 82, CH-4056 Basel, Switzerland}

\date{\today}

\begin{abstract}
A single photon source is a key enabling technology in device-independent quantum communication \cite{Barrett2005,Acin2007}, quantum simulation for instance boson sampling \cite{Aaronson2011,WangPRL2019}, linear optics-based  \cite{Rudolph2017,Wang2019review} and measurement-based quantum computing \cite{Raussendorf2007}. These applications involve many photons and therefore place stringent requirements on the efficiency of single photon creation. The scaling on efficiency is an exponential function of the number of photons. Schemes taking full advantage of quantum superpositions also depend sensitively on the coherence of the photons, i.e.\ their indistinguishability \cite{Sangouard2012}. It is therefore crucial to maintain the coherence over long strings of photons. Here, we report a single photon source with an especially high system efficiency: a photon is created on-demand at the output of the final optical fibre with a probability of 57\%. The coherence of the photons is very high and is maintained over a stream consisting of thousands of photons; the repetition rate is in the GHz regime. We break with the established semiconductor paradigms, such as micropillars  \cite{Somaschi2016,Loredo2016,Ding2016,Wang2016,Unsleber2016,Wang2019}, photonic crystal cavities \cite{Feng2018} and waveguides \cite{Acari2014,Thyrrestrup2018,Uppu2020}. Instead, we employ gated quantum dots in an open, tunable microcavity.
The gating ensures low-noise operation; the tunability compensates for the lack of control in quantum dot position and emission frequency; the output is very well-matched to a single-mode fibre. An increase in efficiency over the state-of-the-art by more than a factor of two, as reported here, will result in an enormous decrease in run-times, by a factor of $10^{7}$ for 20 photons \cite{WangPRL2019}.
\end{abstract}

\maketitle

\normalsize

A single emitter can be used as a single photon source. Unlike a cold atom in vacuum \cite{McKeever2004}, an emitter in the solid-state is naturally trapped in space \cite{Aharonovich2016}. Semiconductor quantum dots have large optical dipole moments, very high radiative efficiency and a relatively weak coupling to phonons, advantages over other solid-state emitters \cite{Kurtsiefer2000,Babinec2010,Sipahigil2014}. A single quantum dot under resonant excitation at low temperature mimics a two-level system. Photonic engineering on a nano- or micro-scale is required to funnel the photons into one specific mode and to couple the photons from this mode into a single-mode fibre \cite{Lodahl2015,Senellart2017}. There are two established techniques. First, in a resonant microcavity, photons are emitted preferentially into the microcavity mode (the Purcell effect), and in an asymmetric microcavity, photon leakage from the microcavity acts as an out-coupler. Much success has been achieved with micropillars \cite{Somaschi2016,Loredo2016,Ding2016,Wang2016,Unsleber2016,Wang2019} (for which the $\beta$-factor, the probability of emission into the microcavity mode, is as high as 88\% \cite{Somaschi2016}) and with photonic crystal cavities \cite{Feng2018}. Secondly, in an on-chip waveguide, photons are emitted preferentially into a laterally-propagating mode and a grating couples the light off the chip \cite{Acari2014,Thyrrestrup2018,Uppu2020}. In this case, a $\beta$-factor as high as 98\% has been demonstrated \cite{Acari2014}. However, in both schemes, the system efficiency, the efficiency as measured in the final optical fibre, is limited by losses and inefficiencies in the out-coupling. The highest system efficiency reported to date is 24\% \cite{Wang2019}.

Coherence depends sensitively on the noise in the device. Charge noise results in a fluctuating emission frequency; it may also result in telegraph noise should the charge state of the quantum dot itself fluctuate. Charge noise is extremely low in gated, high-quality material \cite{Kuhlmann2013}. In particular, the charge state of the quantum dot can be locked by Coulomb blockade \cite{Warburton2000}. This not only eliminates telegraph noise associated with a fluctuating quantum dot charge but also allows a single electron (or hole) to be trapped on the quantum dot, facilitating a spin-photon interface. In a ``best of both worlds" device, photonic engineering is combined with gated operation.

Here, we take the microcavity-route to generating single photons from single quantum dots. We use an open microcavity \cite{Barbour2011,Najer2019,Wang2019NP}. There are some compelling features: the microcavity is tunable \cite{Barbour2011}; the output is very close to a simple Gaussian mode; it is straightforward to incorporate gates \cite{Najer2019}; scattering and absorption losses are extremely small \cite{Najer2019}. 

In the generic case (Jaynes-Cummings Hamiltonian with atom-cavity coupling $g$, cavity loss-rate $\kappa$, atom decay rate into non-cavity modes $\gamma$), $\beta=F_{\rm P}/(F_{\rm P}+1)$ where the Purcell factor is $F_{\rm P}=4g^2/(\kappa \gamma)$. The conversion efficiency of an exciton in the quantum dot to a photon exciting the cavity is $\eta = \beta \cdot \kappa/(\kappa+\gamma)$. For fixed $g$ and $\gamma$, $\eta$ can be maximised \cite{Cui2005} by choosing $\kappa=2g$. Taking a quantum dot with transform-limited linewidth ($\gamma/(2\pi)=0.29$ GHz) in a microcavity ($g/(2\pi)=4.4$ GHz) \cite{Najer2019}, the condition $\kappa=2g$ implies an efficiency $\eta$ as high as 94\%. In other words, ideal behaviour results in high efficiency single photon generation. We present a realisation of this concept.

\begin{figure*}[t!]
\centering
\includegraphics[width=\textwidth]{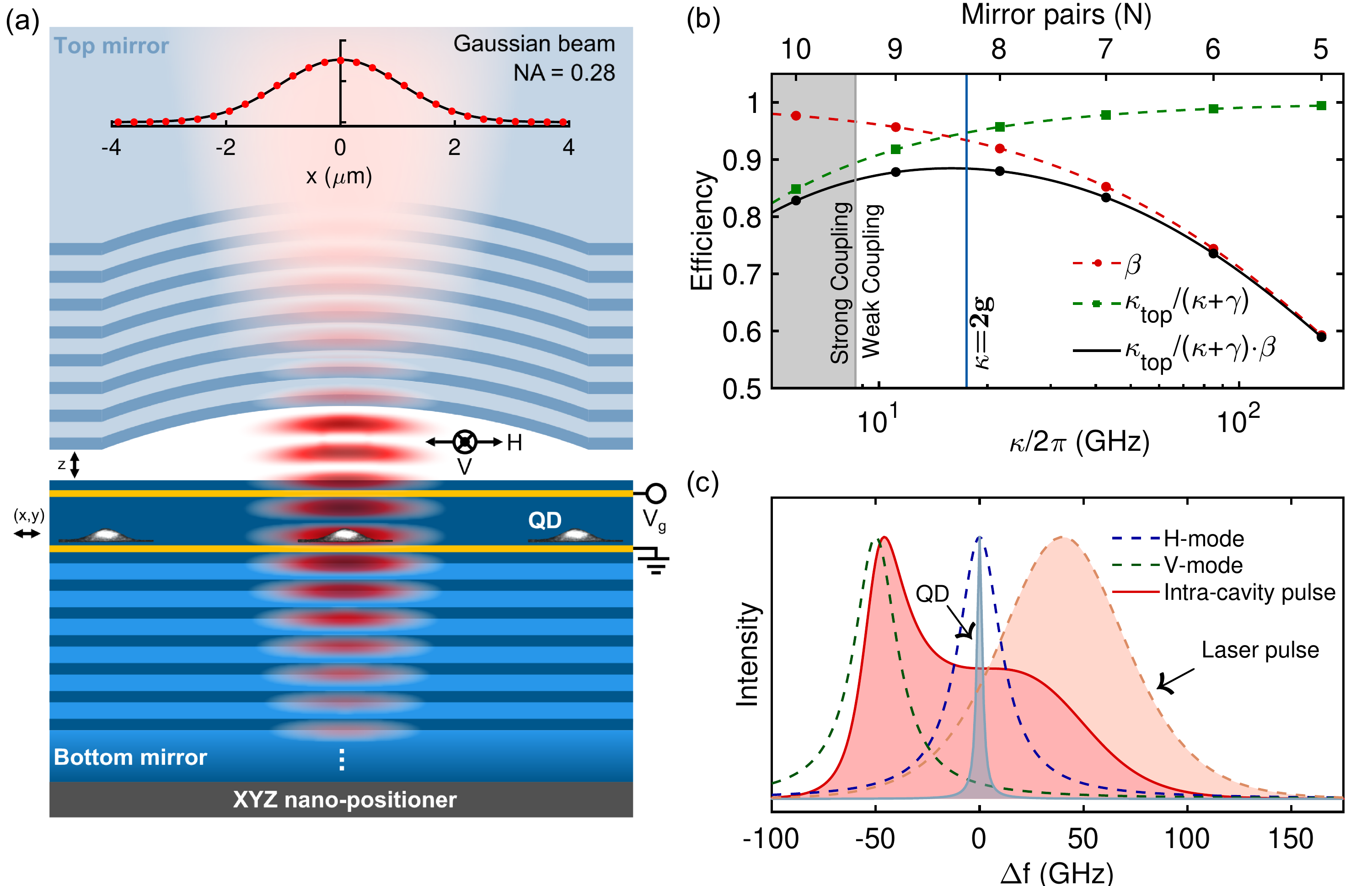}
\caption{(a) The microcavity. The semiconductor heterostructure consists of a GaAs/AlAs Bragg mirror, the ``bottom" mirror, and a p-i-n diode. The InGaAs quantum dots are located in the intrinsic region, in tunnel-contact with the Fermi sea in the n-layer. The position of the heterostructure can be adjusted ($\leftrightarrow$, $\updownarrow$) with respect to the ``top" mirror, a concave mirror in a silica substrate, using an XYZ-nanopositioner. A simulation (red points) shows that the output is very close to a Gaussian beam (black line). (b) Calculated conversion efficiency, quantum dot exciton to photon exiting the top mirror,  as a function of the microcavity decay rate $\kappa$ for ``atom"-photon coupling $g/(2\pi)=4.3$ GHz and atom decay rate $\gamma/(2\pi)=0.30$ GHz. $\eta=\kappa_{\rm top}/(\kappa+\gamma) \cdot \beta$; $\beta=F_{\rm P}/(F_{\rm P}+1)$ with $F_{\rm P}=4g^2/(\kappa \gamma)$. (c) Excitation scheme. The quantum dot is in resonance with the H-polarised microcavity mode; the laser is blue-detuned and V-polarised. The driving intensity as experienced by the quantum dot is shown.}
\end{figure*}

We use a highly miniaturised Fabry-Perot cavity (Fig.~1a). The ``top" mirror has a concave shape and is micro-machined into a silica substrate (Supplementary Information Sec.~II); the ``bottom" mirror is a highly reflective planar mirror, part of the semiconductor heterostructure \cite{Barbour2011} (Supplementary Information Sec.~I). Quantum dots in this structure exhibit close-to-transform-limited linewidths \cite{Najer2019}. With a highly reflective top mirror, the microcavity has $\mathcal{Q}$-factors up to $10^{6}$ and the strong coupling regime of cavity-QED can be reached \cite{Najer2019}. This allows a precise measurement of the coupling ($g/(2\pi)=4.0$ GHz) and an estimation of the residual losses in the semiconductor (373 ppm per round-trip). Here, we use a modest reflectivity top mirror (transmission 10,300 ppm per round-trip according to the design) chosen such that $\kappa \simeq \kappa_{\rm top} \gg \kappa_{\rm bottom}$ and $\kappa \approx 2g$  (Fig.~1b). The measured $\mathcal{Q}$-factor is 12,600, matching the value expected from the design of the two mirrors (Supplementary Information Sec.~V). At these $\mathcal{Q}$-factors, the residual losses in the semiconductor are negligible. The heterostructure contains thin n- and p-type layers with the quantum dots in tunnel contact with the electron Fermi sea in the n-type layer such that Coulomb blockade is established (Supplementary Information Sec.~I). It is straightforward to make contacts to the n- and p-type layers even in the full microcavity structure. The chip is positioned relative to the top mirror {\it in situ} (Fig.~1a): this tunability is exploited to ensure a match between a particular quantum dot and the microcavity mode, both in frequency and lateral position.

\begin{figure*}[t!]
\centering
\includegraphics[width=\textwidth]{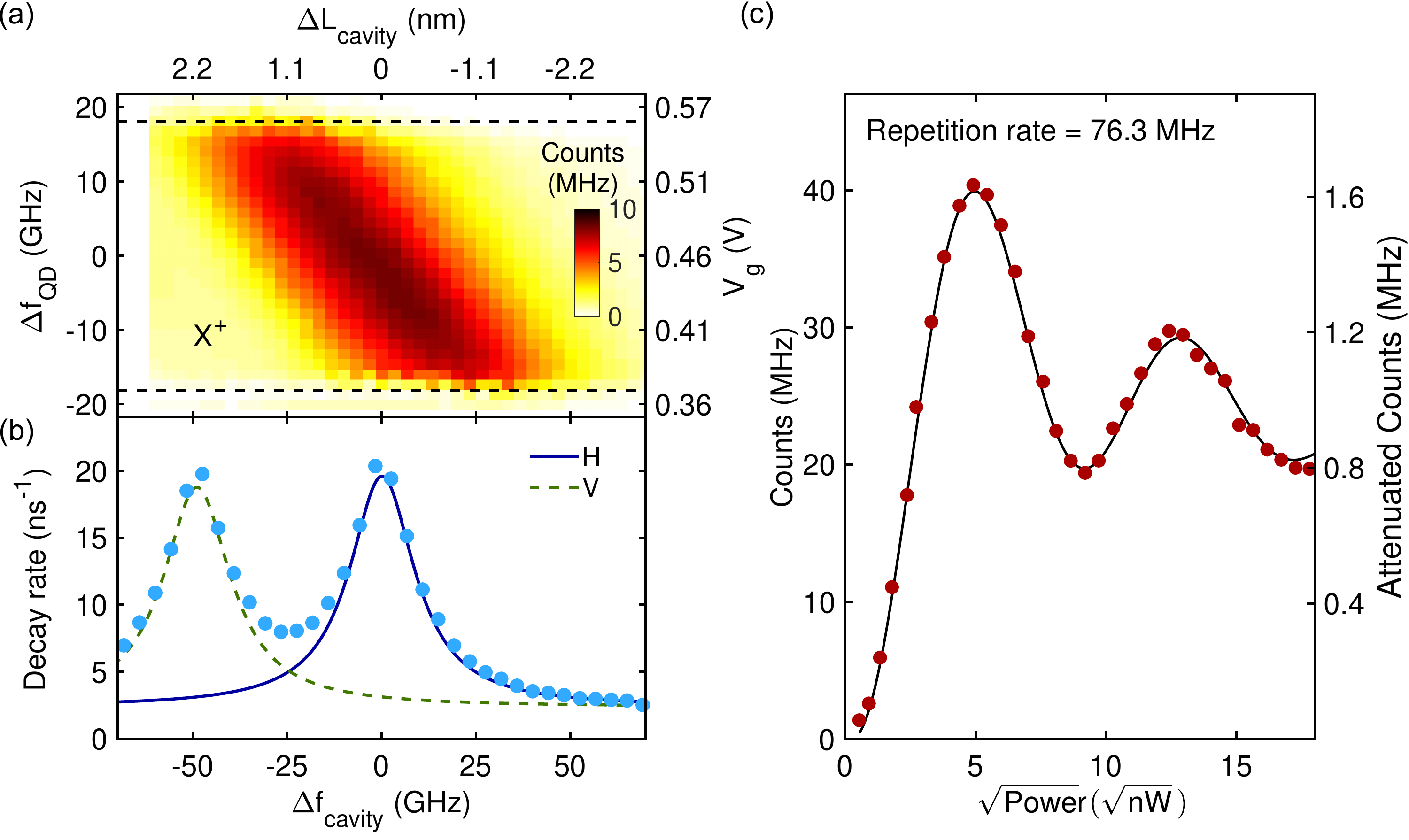}
\caption{Single photon flux. (a) Quantum dot (QD1) signal versus z-piezo voltage (microcavity detuning) and bias voltage (quantum dot detuning). The positively charged trion X$^{+}$ is resonant with the microcavity; the dashed lines denote the boundaries of the Coulomb blockade plateau. (b) Radiative decay rate $\gamma$ (following pulsed resonant excitation) versus microcavity detuning for constant bias and constant $(x,y)$-position. The total Purcell factor $F_{\rm P}$ is determined to be 11 implying $\beta=93$\%. Via Lorentzian fits, the $\beta$-factor specific to the H-polarised mode is determined: $\beta_{\rm H}=86$\%. (c) Measured signal versus square root of laser power for zero microcavity-X$^{+}$ detuning. The laser repetition frequency is 76.3 MHz; the detector has an efficiency of $42\pm3$\%. The signal is deliberately attenuated by a factor of 9.9 (right $y$-axis). The left $y$-axis shows the expected signal without the attenuation and with a perfect detector. The solid-line is the result of the calculation, describing the response of the quantum dot to the driving field (Fig.~1c).}
\end{figure*}

A challenge in all optically-driven quantum dot single photon sources is to separate the single photon output from the driving laser light. A standard scheme is to excite and detect in a cross-polarised configuration \cite{KuhlmannRSI2013}. Applied to a charged exciton for which the transitions are circularly polarised, this scheme leads to a 50\% loss in the collection efficiency. Here, we avoid this loss. We work with the positively-charged exciton, X$^{+}$. The microcavity mode splits into two modes, H- and V-polarised, separated by 50 GHz, on account of a small birefringence (Supplementary Information Sec.~III); the spectrum of the laser pulses is larger than this splitting (Fig.~1c). The quantum dot is tuned into resonance with the higher-frequency, H-polarised mode. The laser is V-polarised and blue-detuned with respect to both microcavity modes such that the tails of the laser spectrum and the V-polarised microcavity mode overlap at the frequency of the H-polarised mode (Fig.~1c). The quantum dot emits preferentially into the H-polarised microcavity mode. A cross-polarised scheme now separates the V-polarised laser pulses from the H-polarised single photons with a loss depending only on the unwanted coupling of the quantum dot to the V-polarised mode (Supplementary Information Sec.~VI). Provided that the mode-splitting is larger than the mode linewidths \cite{Wang2019}, this loss is small.

Experimentally, we choose a quantum dot and maximise the coupling of the X$^{+}$-resonance to the microcavity. To do this, we record a decay curve following resonant excitation: the radiative decay rate is largest at maximum coupling. The quantum dot and microcavity frequencies are tuned to establish a resonance (Fig.~2a). The Purcell-factor is determined by scanning the microcavity frequency: on resonance with a microcavity mode, the decay time is just 47.5 ps; far detuned, the decay time tends to 520 ps, resulting in $F_{\rm P}=10$ for QD1 (12 for QD6) (Fig.~2b). On resonance with the H-polarised microcavity mode, we determine $\beta_{\rm H}$, the probability of emission into the H-polarised mode, to be $\beta_{\rm H}=86$\%.

We now maximise the flux of single photons. Implementing the excitation scheme (Fig.~1c), the central frequency of the laser is tuned to find the maximum signal. As a function of laser power, the quantum dot signal exhibits oscillations, indicative of Rabi oscillations (Fig.~2c). The laser power is set at the maximum signal corresponding to the best implementation of a $\pi$-pulse. An intensity autocorrelation measurement demonstrates clear photon antibunching and a high purity of single photon generation, $g^{(2)}(0)=2.1$\% (Fig.~3a). The purity is limited by a small amount of laser light leaking into the detection channel (0.3\% of total signal) and double-excitation events.

The main new feature over previous experiments is the very high efficiency of the source. On excitation with a $\pi$-pulse, we obtain an on-demand, coherent single photon at the output of the collection fibre (a standard optical fibre) with a probability of 57\%. The efficiency is determined from the photon flux. At a repetition frequency of 76.3 MHz, we attenuate the beam by a factor of 9.9 (to avoid saturating the detector) and measure the count rate (Fig.~2c). Taking account of the detector efficiency and a small nonlinearity in the detector's response (Supplementary Information Sec.~VII), we determine the system efficiency, the probability of creating a single photon at the output of the system's final optical fibre, to be $53 \pm 3$\% for QD1 ($57 \pm 3$\% for QD6). 

\begin{figure}[b!]
\includegraphics[width=0.5\textwidth]{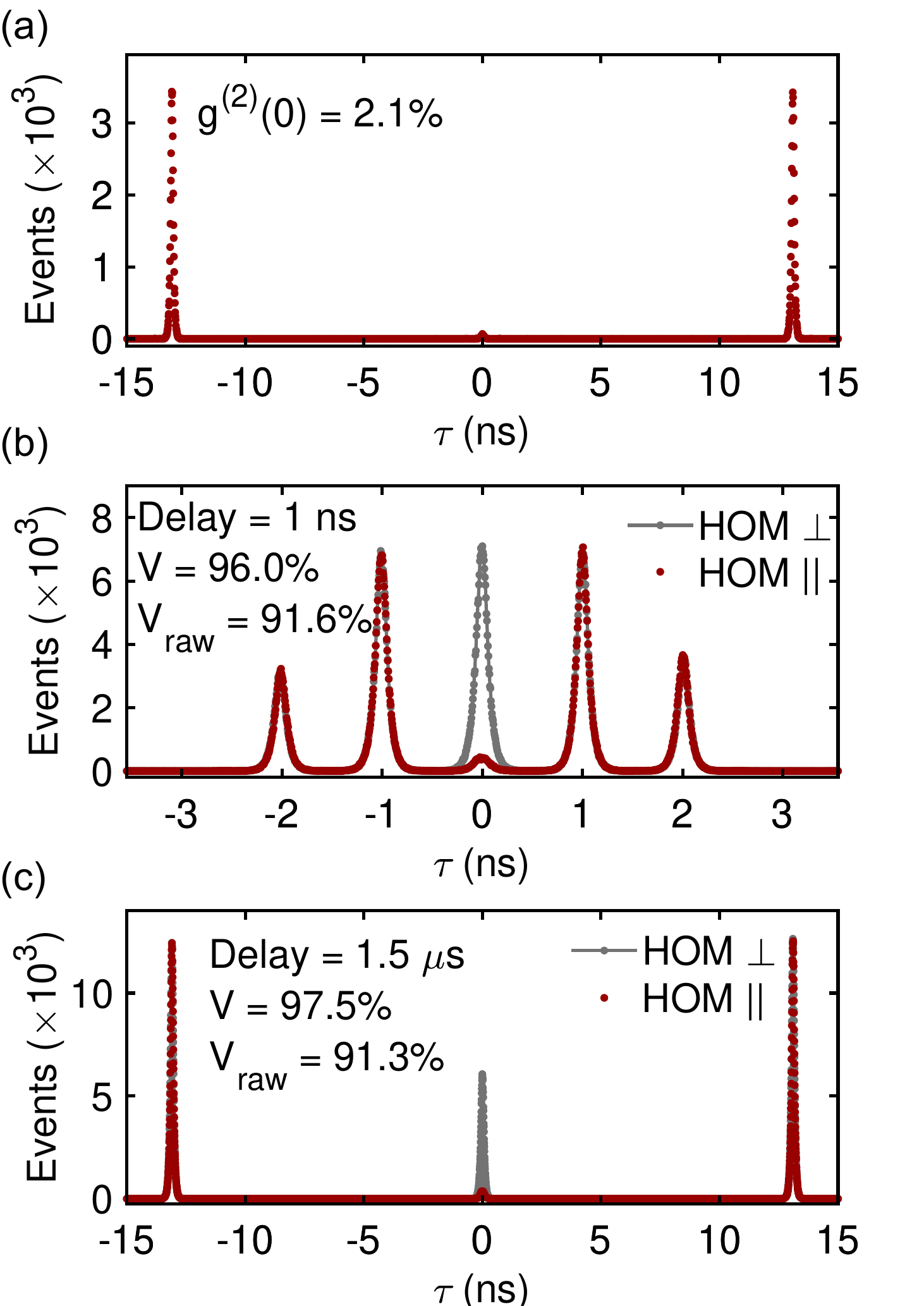}
\caption{Quantum-optics characterisation. (a) Autocorrelation $g^{(2)}$ versus delay $\tau$ (QD1). (b), (c) Hong-Ou-Mandel (HOM) experiment (QD1) showing two-photon interference for photons created 1 ns and 1.5 $\mu$s apart in time, (b) and (c), respectively.}
\end{figure} 

The coherence of the single photons is probed with two-photon interference, a Hong-Ou-Mandel (HOM) experiment. On creating two photons 1 ns apart in time, the raw HOM visibility is $V_{\rm raw}=91.6$\% (Fig.~3b). Correcting for a small imperfection in the HOM interferometer, $V_{\rm raw}=92.5$\%. The HOM visibility is negatively influenced by the finite $g^{(2)}(0)$: following the standard procedure \cite{Santori2002}, the ``true" photon overlap can be estimated to be $V \simeq (1+2 g^{(2)}(0)) \cdot V_{\rm raw} = 96.7$\% (Supplementary Information Sec.~IV). This demonstrates that successively generated photons are highly coherent. Crucial however is the coherence of photons separated much further apart in time. The HOM visibility on interfering two photons separated by 1.5 $\mu$s in time is equally high (Fig.~3c). Given that photons can be created each nano-second (Fig.~3b), these experiments demonstrate that the device can produce a string consisting of thousands of highly coherent photons. The coherence time of the source is clearly much larger than 1.5 $\mu$s (Supplementary Information Sec.~IV).

The single-photon source is very stable in time. The noise in the single photon flux is limited by shot-noise on time-scales of one hour (Fig.~4a), increasing only slightly on time-scales of multiple hours (Fig.~4b). The tunability of the microcavity enables us to bring multiple quantum dots one-by-one into resonance with the same microcavity mode. Six quantum dots were investigated in detail. All six have essentially the same values of single photon purity, system efficiency (Fig.~4c) and coherence (Fig.~4d).

\begin{figure*}[t!]
\centering
\includegraphics[width=\textwidth]{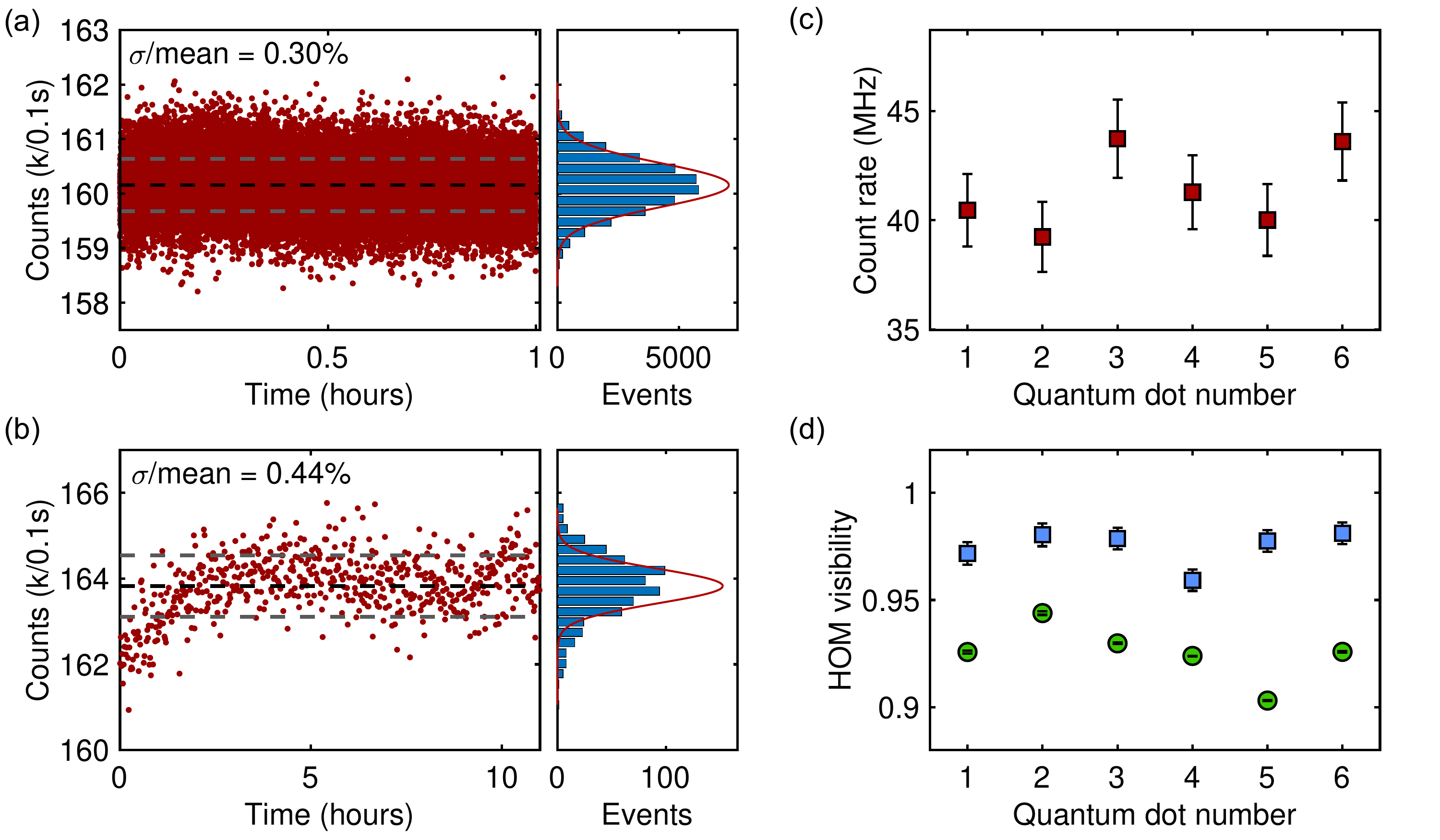}
\caption{Stability. Single photon flux versus time and associated histogram recorded over one hour and over ten hours, (a) and (b), respectively, on QD1. Maximum count rates and HOM visibilities recorded on six separate quantum dots, (c) and (d), respectively. The blue data points correspond to the corrected visibility of the source, $V$; the green points represent $V_{\rm raw}$.}
\end{figure*} 

The system efficiency, $\Sigma$, here 53\%--57\% (QD1--QD6), is a product of factors, $\Sigma=\pi  \cdot \beta_{\rm H} \cdot \kappa_{\rm top}/(\gamma + \kappa_{\rm total}) \cdot \eta_{\rm optics}$ where $\pi$ is the probability of producing a photon on excitation with a laser pulse; and $\eta_{\rm optics}$ represents the throughput of the entire optical system (from microcavity to the output from the final output fibre). $\beta_{\rm H}$ and $\kappa_{\rm top}/(\gamma + \kappa_{\rm total})$ are both determined precisely in the experiment, 86\% and 96\%, respectively. $\beta_{\rm H}$ matches theoretical expectations based on the optical dipole moment and the microcavity geomertry (Supplementary Information Sec.~V). To determine $\pi$, we describe the excitation scheme, a detuned laser pulse followed by ring-down, and how it drives a two-level system, including an intensity-dependent phonon-related dephasing process \cite{Ramsay2010} (Supplementary Information Sec.~VIII). This calculation describes the Rabi oscillations very successfully (Fig.~2c), enabling us to infer that at the peak signal, $\pi=96.3$\%. The remaining factor, $\eta_{\rm optics}$ (68\%), accounts for the throughput losses in the optical components, losses on coupling the single photons into the single-mode fibre, and reflection losses at three surfaces (upper surface of top mirror, two fibre facets) which lacked an antireflection coating. The conclusion of this analysis is that the main contribution to the losses lies in $\eta_{\rm optics}$, i.e.\ in the classical optics. The quantum effects, the coupling to the vacuum mode of the microcavity and the scheme to invert the two-level system via the off-resonance microcavity mode, are under excellent control.

Based on this analysis, a single photon source with a system efficiency of 80\% is within reach by eliminating the losses in the optical components. Even better performance is conceivable by improving the quantum effects -- an increased coupling (via miniaturisation of the top mirror), a decreased bare decay rate (via lateral structuring), and a larger microcavity mode-splitting will all increase the efficiency. In addition to the applications as a single photon source, a further broad area of application exploits the spin of the trapped hole. Implementing spin manipulation in the microcavity device, for instance via lateral excitation (an ``atom" drive), will facilitate applications such as a single photon transistor \cite{Chang2007}, the efficient and fast creation of spin-photon entangled pairs, and an efficient source of multi-photon cluster states \cite{Schwartz2016}.\\
 
\noindent {\bf Acknowledgements} We thank Alistair Brash, Peter Lodahl, Nicolas Sangouard and Sebastian Starosielec for fruitful discussions. We acknowledge financial support from SNF project 200020\_156637, NCCR QSIT and Horizon-2020 FET-Open Project QLUSTER. A.J.\ acknowledges support from the European Unions Horizon 2020 Research and Innovation Programme under the Marie Sk{\l}odowska-Curie grant agreement No.\ 840453 (HiFig). S.R.V., R.S., A.L.\ and A.D.W.\ acknowledge gratefully support from DFH/UFA CDFA05-06, DFG TRR160, DFG project 383065199, and BMBF Q.Link.X.\\

\noindent {\bf Author contributions} N.T.\ and A.J.\ contributed equally to this work. N.T., A.J., N.O.A.\ and D.N.\ carried out the microcavity experiments. M.C.L.\ characterised the quantum dots and optimised the photon counting hardware. N.T.\ and D.N.\ fabricated the curved mirror. D.N., A.L.\ and R.J.W.\ designed the heterostructure. D.N.\ developed the surface passivation technique. A.R.K., R.S., S.R.V., A.D.W.\ and A.L.\ fabricated the semiconductor device. A.J.\ developed the model of the excitation mechanism. D.N.\ carried out the numerical simulations of the microcavity mode. N.T., A.J.\ and R.J.W.\ wrote the paper with input from all authors.\\

\noindent {\bf Competing interests} The authors declare no competing interests.\\

\noindent {\bf Correspondence and requests for materials} should be addressed to A.J.\\

\newpage

\renewcommand{\figurename}{Supplementary Fig.}
\setcounter{figure}{0}

\section*{Supplementary Information}

\section{Heterostructure design and growth}

\begin{figure*}[t!]
	\centering
	\includegraphics[width=160 mm]{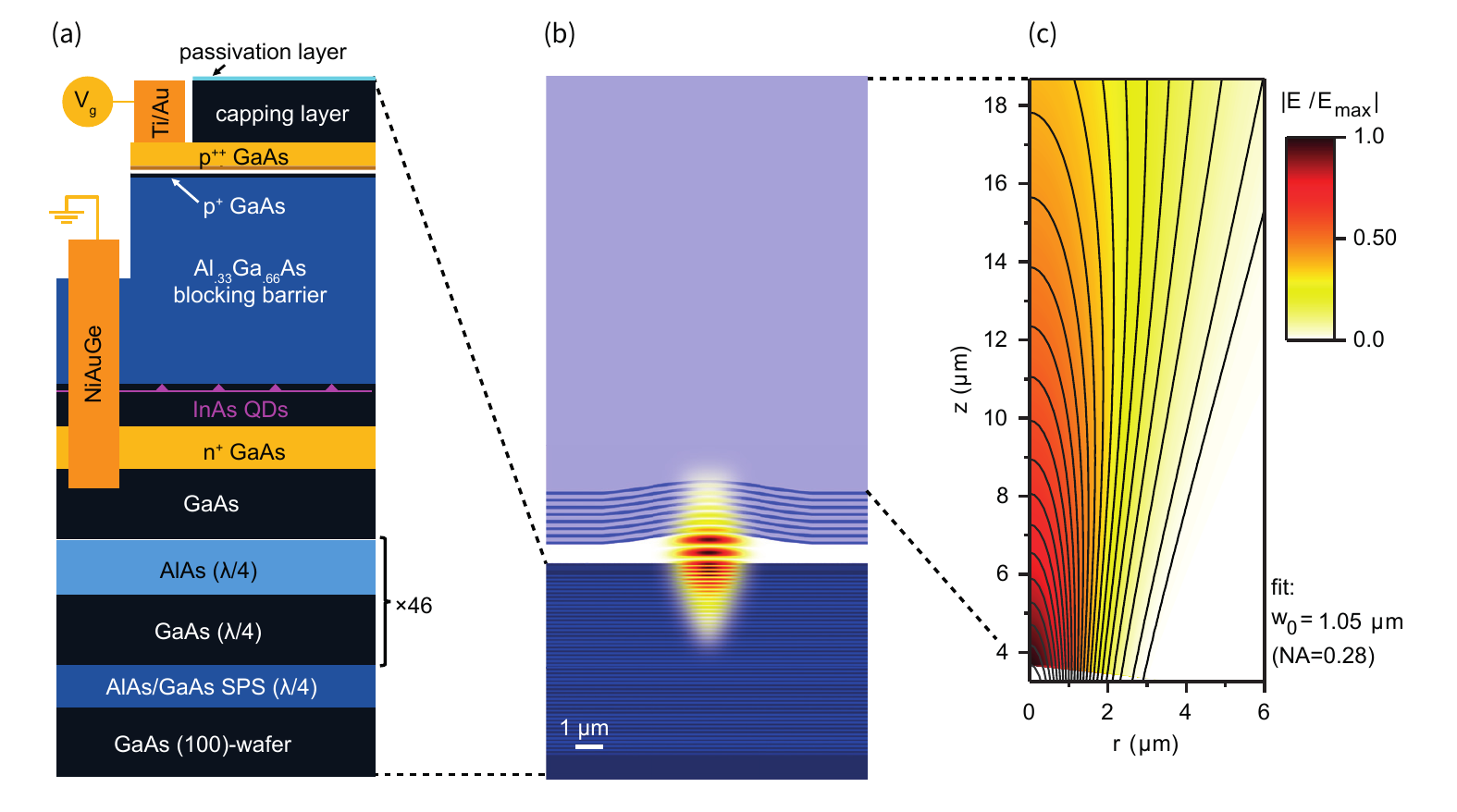}
	\begin{singlespace}
 \caption{\footnotesize
 \textbf{Heterostructure design and numerical simulation of the microcavity.}
 \textnormal{{\bf (a)} The semiconductor heterostructure consists of a DBR and an n-i-p diode structure with embedded self-assembled InGaAs QDs. {\bf (b)} Numerical simulation of the vacuum electric field $|E_{\rm vac}|$ confined by the microcavity (image to scale). {\bf (c)} Colour-scale plot: normalised electric field within the SiO$_2$ substrate supporting the ``top'' mirror. Contour lines: fit of a Gaussian beam to the calculated normalised electric field. The fit yields a beam waist of $w_0=1.05$\,$\mu$m corresponding to a numerical aperture of ${\rm NA}=0.279$. $|E_{\rm max}|$ is the maximum electric field amplitude in this particular domain.}
 }
 \label{fig:mode}
 \end{singlespace}
\end{figure*}

The heterostructure is grown by molecular beam epitaxy (MBE) and consists of an n-i-p diode with embedded self-assembled InGaAs quantum dots (QDs). This design allows for QD frequency tuning via the dc Stark effect as well as QD charging via Coulomb blockade. The n-i-p diode is grown on top of a semiconductor distributed Bragg reflector (DBR), a planar ``bottom" mirror, composed of 46 pairs of AlAs (80.6\,nm thick)/GaAs (67.9\,nm thick) quarter-wave layers (QWLs) with a centre wavelength of nominally 940\,nm (measured: 917\,nm). Below the DBR, an AlAs/GaAs short-period superlattice (SPS) composed of 18 periods of 2.0\,nm AlAs and 2.0\,nm GaAs is grown for stress-relief and surface-smoothing.

From bottom to top (see Supplementary Fig.~\ref{fig:mode}a), the diode consists of an n-contact, 41.0\,nm Si-doped GaAs, n$^+$, doping concentration $2\cdot10^{18}$\,cm$^{-3}$. A 25.0\,nm layer of undoped GaAs acts as a tunnel barrier between the n-contact and the QDs. The self-assembled InGaAs QDs are grown by the Stranski-Krastanov process and the QD emission is blue-shifted via a flushing-step~\cite{Wasilewski1999}. The QDs are capped by an 8.0\,nm layer of GaAs. A blocking barrier, 190.4\,nm of Al$_{.33}$Ga$_{.67}$As, reduces current flowing across the diode in forward-bias. The p-contact consists of 5.0\,nm of C-doped GaAs, p$^+$ (doping concentration $2\cdot10^{18}$\,cm$^{-3}$) followed by 20.0\,nm of p$^{++}$-GaAs (doping concentration $1\cdot10^{19}$\,cm$^{-3}$). Finally, there is a 54.6\,nm-thick GaAs capping layer.

The layer thicknesses are chosen to position the QDs at an antinode of the vacuum electric field. The p-contact is centred around a node of the vacuum electric field to minimise free-carrier absorption in the p-doped GaAs. Coulomb blockade is established on times comparable to the radiative decay time for GaAs tunnel barriers typically $\lesssim 40$\,nm thick. This is less than the thickness of a QWL thereby preventing the n-contact being positioned likewise at a node of the vacuum electric field. However, at a photon energy 200 meV below the bandgap~\cite{Casey1975}, the free-carrier absorption of n$^+$-GaAs ($\alpha\approx10$\,cm$^{-1}$) is almost an order-of-magnitude smaller than that of p$^{++}$-GaAs ($\alpha\approx70$\,cm$^{-1}$). The weak free-carrier absorption of n$^+$-GaAs is exploited in the design presented here by using a standard 25\,nm thick tunnel barrier. The n-contact is positioned close to a vacuum field node, although not centred around the node itself.

After growth, individual $3.0 \times 2.5$\,mm$^2$ pieces are cleaved from the wafer. The QD density increases from zero to $\sim 10^{10}$ cm$^{-2}$ in a roughly centimetre-wide stripe across the wafer. The sample used in the experiments presented here was taken from this stripe. Its QD density, measured by photoluminescence imaging, is approximately $7 \times 10^{6}$ cm$^{-2}$. 

Separate ohmic contacts are made to the p$^{++}$ and n$^{+}$  layers. For the n-contact, the capping layer, the p-doped layers and part of the blocking barrier are removed by a local etch in citric acid. On the new surface, NiAuGe is deposited by electron-beam physical vapour deposition (EBPVD). Low-resistance contacts form on thermal annealing. To contact the p-doped layer, the capping layer is removed by another local etch. On the new surface, a Ti/Au contact pad (100\,nm thick) is deposited by EBPVD. Although this contact is not thermally annealed, it provides a reasonably low-resistance contact to the top-gate on account of the very high p-doping (Supplementary Fig.~\ref{fig:mode}a).

After fabricating the contacts to the n- and p-layers, the contacts are covered with photoresist and a passivation layer is deposited onto the sample surface. A thin native oxide layer on the surface is removed by etching a few nm of GaAs in HCl. Following a rinse in deionised water, the sample is immersed in a bath of ammonium sulphide ((NH$_4$)$_2$S). Subsequently, the sample is transferred rapidly into the chamber of an atomic-layer deposition (ALD) setup. An 8\,nm layer of Al$_2$O$_3$ is deposited using ALD at a temperature of 150$^\circ$C. With the present heterostructure, this process is essential to reduce surface-related absorption: a low-loss microcavity is only achieved following surface-passivation~\cite{Najer2019}. An advantage of the surface passivation lies in the fact that it prevents the native oxide of GaAs from re-forming after its removal: it provides a stable termination to the GaAs heterostructure~\cite{Guha2017}. Following the surface-passivation procedure and photoresist stripping, the NiAuGe and Ti/Au films are wire-bonded to large Au pads on a sample holder. Using silver paint, macroscopic wires (twisted pairs) are connected to the Au pads.

When applying a voltage across the gates of this n-type device, the neutral exciton, X$^0$, is observed at intermediate biases. The negatively-charged trion, X$^-$, is observed at more positive bias, and the positively-charged trion, X$^+$ at more negative voltages. This particular device presents a small leakage current at the X$^-$ voltage, making it more appropriate to work with X$^+$ instead. Given a source of holes, n-type devices exhibit Coulomb blockade of positively-charged excitons~\cite{Ediger2007}. The X$^0$ and X$^+$ can both be excited by the same laser pulse. The splitting between X$^0$ and X$^+$ varies from QD to QD in a range between 606 GHz (QD1) to 143 GHz (QD3). At lower voltages, the QD is initially empty. On exciting an X$^0$ the electron tunnels out rapidly leaving a single hole, allowing the X$^+$ transition. Should the QD lose its residual hole for any reason, the process repeats very rapidly.

\section{Curved mirror fabrication}
\label{sec:craterfab}
The top mirror is fabricated in a 0.5\,mm thick fused-silica substrate. An atomically-smooth crater is machined at the silica surface via CO$_2$-laser ablation \cite{Hunger2012,Greuter2014}. We achieve craters with a similar radius of curvature as described in Ref.~\cite{Greuter2014}, but with a shallower profile by substituting the focusing lens in the ablation setup by a lens with ${\rm NA}=0.67$.

The profile of the fabricated crater is measured by a confocal laser scanning microscope (Keyence Corporation), as shown in Supplementary Fig.~\ref{fig:crater}a. From the two-dimensional height profile, two principal axes can be identified, and the profile parameters can be extracted (Supplementary Fig.~\ref{fig:crater}b). The radius of curvature of this crater is $R=(11.98 \pm 0.02)\,\mu$m and the sagittal height $s = (0.41 \pm 0.02)\, \mu$m. After laser ablation, the crater is coated with 8 QWL-pairs of Ta$_2$O$_5$ (refractive index $n=2.09$ at $\lambda_0=920$\,nm) and SiO$_2$ ($n=1.48$ at $\lambda_0=920$\,nm) layers (terminating with a layer of Ta$_2$O$_5$) by ion-beam sputtering at a commercial company (Laseroptik GmbH), see Supplementary Fig.~\ref{fig:mode}b.

\begin{figure*}[t!]
	\centering
	\includegraphics[width=160 mm]{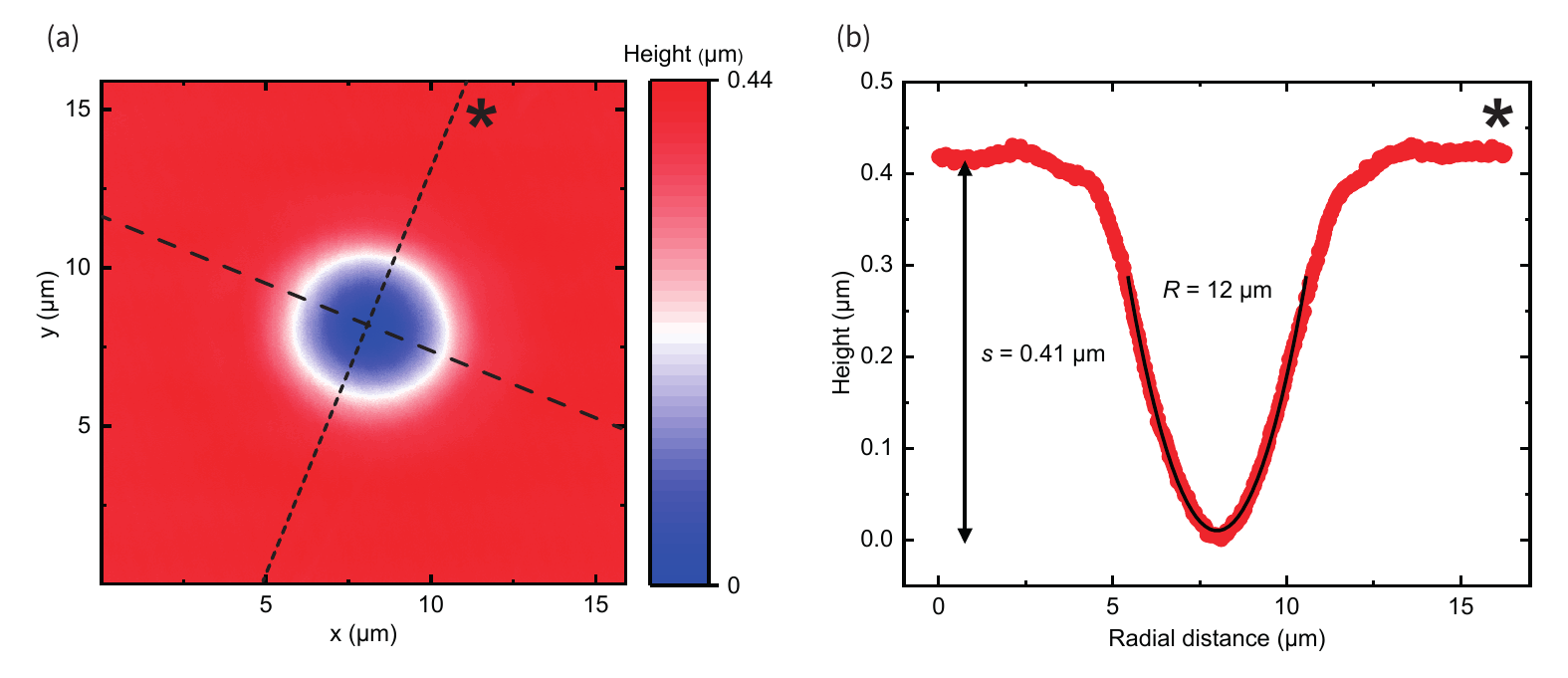}
	\begin{singlespace}
	\caption{\footnotesize
	\textbf{Geometrical characterisation of the curved mirror.}
	\textnormal{Following CO$_2$-laser machining, the fabricated crater's profile is measured with a confocal laser scanning microscope. \textbf{(a)} Height map of the crater determined with sub-nm resolution. From the height map, the two principal planes are extracted by fitting a two-dimensional Gaussian function to the data. \textbf{(b)} By evaluating the height information along the two principal axes, it is possible to extract the crater's parameters such as the radius of curvature $R = (11.98 \pm 0.02)\,\mu$m, sagittal height $s = (0.41 \pm 0.01)\,\mu$m, and asymmetry of 4.5\%. This crater is exactly the one used in the single photon experiments described in the main text.}
 }
 \label{fig:crater}
 \end{singlespace}
\end{figure*}

\section{Microcavity characterisation}
\label{characterisation}
The microcavity is a highly miniaturised Fabry-Perot type resonator. A fundamental mode is resonant for a given laser frequency at a particular microcavity length. In order to determine the $\mathcal{Q}$-factor of the microcavity, a dark-field measurement is performed, as shown in Supplementary Fig.~\ref{fig:cavitymodes}. Given the spectral tunability of the microcavity, its $\mathcal{Q}$-factor can be determined for a wide wavelength range within the stopband of the mirrors, centred around $\lambda_0=919$\,nm.

Supplementary Fig.~\ref{fig:cavitymodes} shows such a measurement performed on a fundamental mode at $\lambda_0=922$\,nm. The fundamental mode splits into two modes, each linearly polarised, with opposite polarisations, H and V. The mode-splitting is 34.6\,GHz in Supplementary Fig.~\ref{fig:cavitymodes}. The H and V axes align with the crystal axes of the semiconductor wafer. This points to the physical origin of the mode-splitting: a small birefringence in the semiconductor. The birefringence is probably induced by a very small uniaxial strain. The splitting of the fundamental microcavity mode into two separate modes together with the linear, orthogonal polarisations of these two modes are exploited in the experiment to achieve high efficiencies in our experiment, as discussed in Sec.~\ref{sec:theory}. The mode-splitting is, therefore, an important parameter. Performing this measurement at different locations on the sample yields similar values of $\mathcal{Q}$-factor but a spread in mode-splittings. For the quantum dots investigated, QD1 to QD6, the splitting lies between 34.6 (QD6) and 50 GHz (QD1).

The $\mathcal{Q}$-factors of both H- and V-polarised modes are extracted from the dark-field spectrum (green and blue curves in Supplementary Fig.~\ref{fig:cavitymodes}) yielding $\mathcal{Q}_{\rm H} = 11,900$ and $\mathcal{Q}_{\rm V} = 12,800$. The finesse is $\mathcal{F} = 506 \pm 13$. $\mathcal{F}$ was determined by microcavity scanning at a wavelength of 922\,nm, the same wavelength used for the determination of the $\mathcal{Q}$-factors. Unlike the mode-splitting, the $\mathcal{Q}$-factors are the same at different locations on the sample.

The microcavity does not have a monolithic design and is potentially susceptible to environmental noise, vibrations and acoustic noise. The microcavity is operated in a helium bath-cryostat. The cryostat is shielded from vibrational noise by an active damping stage and from air-borne acoustic noise by an acoustic enclosure. Using the microcavity itself as a noise sensor shows that environmental noise is significant only when operating with a finesse above $10,000$ \cite{Greuter2014}, corresponding to a $\mathcal{Q}$-factor of approximately $10^{5}$ with the present design. Here, the $\mathcal{Q}$-factor is approximately $10^{4}$ and the experiment was not troubled by residual environmental noise.

\begin{figure}[b!]
	\includegraphics[width=80 mm]{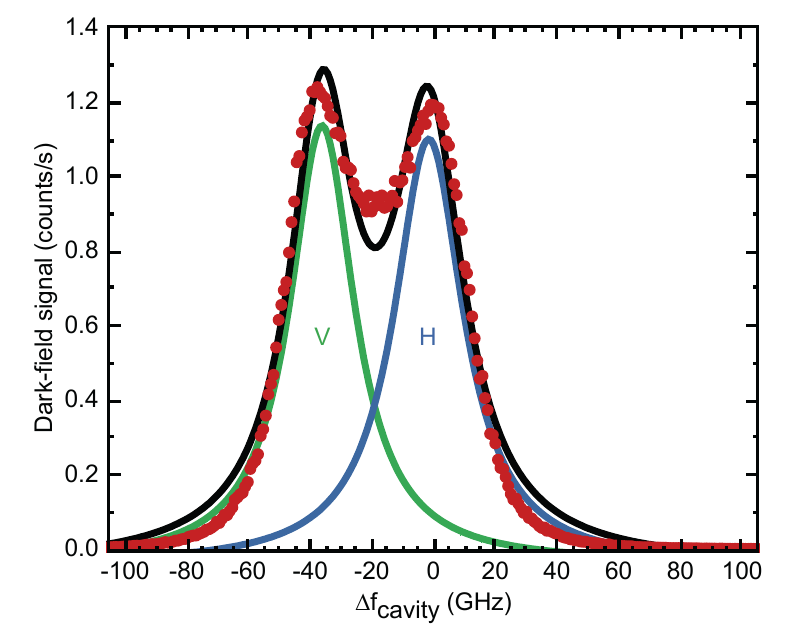}
	\begin{singlespace}
 \caption{\footnotesize
 \textbf{Dark-field spectroscopy of the microcavity.}
 \textnormal{Signal versus optical frequency expressed as a detuning with respect to the upper-frequency resonance. The microscope operates in dark-field mode with principal axes lying at 45 degrees to the principal axes of the microcavity. The wavelength is $\lambda_0=922$\,nm. The fundamental mode splits into two modes both with linear polarisation, one H-polarised, the other V-polarised. The H- and V-axes correspond to the crystal axes of the GaAs wafer. The transmission data (red dots) are fitted to a double-Lorentzian function (blue and green curves) yielding $\mathcal{Q}$-factors for the two polarised modes: $\mathcal{Q}_{\rm H} = 11,900$ and $\mathcal{Q}_{\rm V} = 12,800$. The mode-splitting is 34.6\,GHz.}}
 \label{fig:cavitymodes}
 \end{singlespace}
\end{figure}

\section{Coherence of the single photons}
\begin{figure}[t!]
  \includegraphics[width=80 mm]{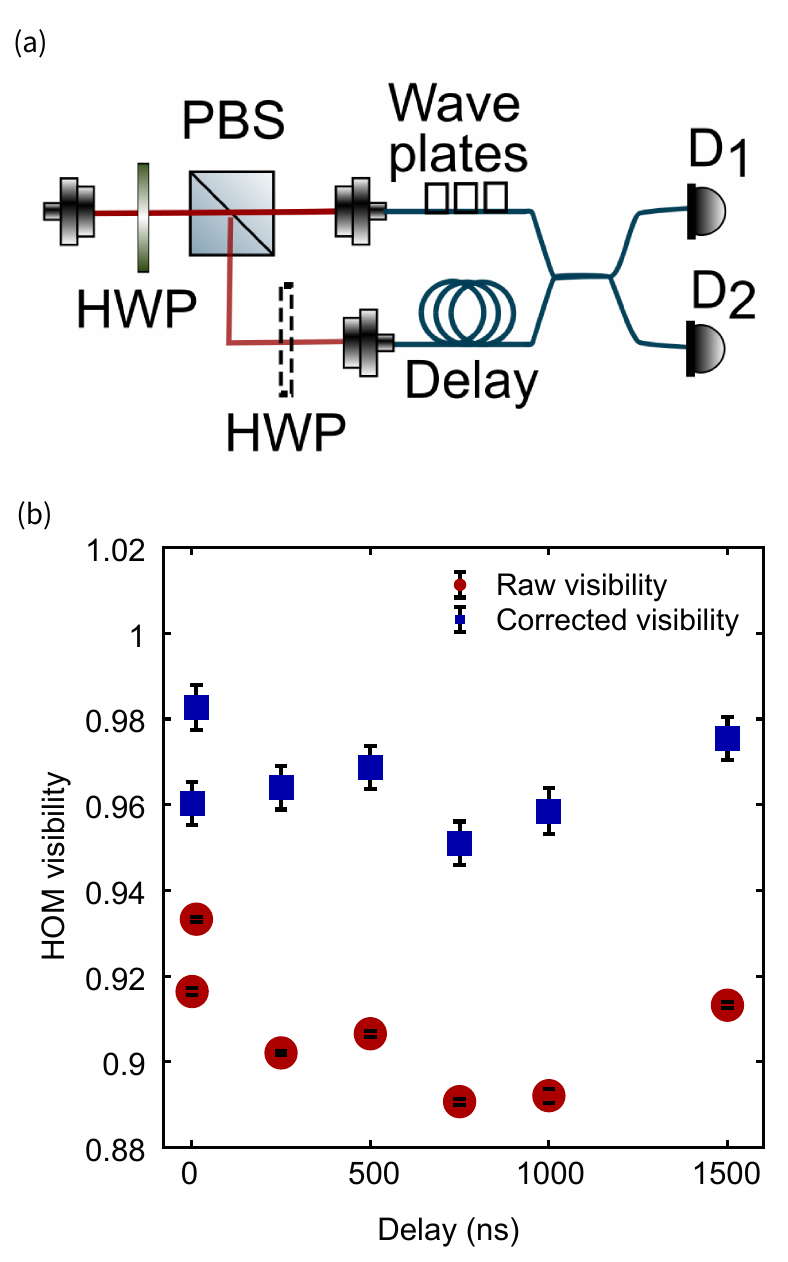}\\
  \caption{\textbf{HOM setup and the visibility of HOM interference versus time delay between the photons.} \textbf{(a)} The optical setup used for HOM measurements. The particular structure of the setup increases the mechanical stability of the interferometer and makes it easy to change the delay between the two photons by changing the fibre delay-loop. \textbf{(b)} $V$ and $V_\mathrm{raw}$ as a function of the delay between the interfering photons.}
\label{fig:HOM}
\end{figure}

We outline the procedure to extract the visibility of the Hong-Ou-Mandel (HOM) interference and present the visibility of the HOM interference as a function of the delay between single photons from the same source. The HOM interference between subsequent photons is measured by launching the stream of single photons into a Mach-Zehnder interferometer with a variable arm. The variable arm introduces a time delay between the photons that interfere. Supplementary Fig.~\ref{fig:HOM}a shows the optical setup for the HOM measurements. The combination of a half-wave plate and a polarising beam splitter (PBS) is used to realise a variable beam-splitter. Three fibre-based wave-retarders are utilised to match the polarisation of the light at the inputs of the fibre beam-splitter, and hence to maximise the classical visibility of the interferometer ($1-\epsilon$). In order to quantify the interference between the two photons, the time delay between the ``clicks" on the two detectors $D_1$ and $D_2$ is measured in the case when the classical visibility of the interferometer is maximised ($\mathrm{HOM}_\parallel$). The red data-points in Fig.~3b and 3c of the main text correspond to $\mathrm{HOM}_\parallel$ measurements. A second half-wave plate can be inserted into the beam path to make the photons from the two arms distinguishable and hence yield the curves labeled $\mathrm{HOM}_{\perp}$, grey data points in Fig.~3b and 3c of the main text. The raw visibility of the HOM interference is calculated as the ratio of the area underneath the curve around zero delay for the two measurements, $V_{\rm raw}=1-\frac{A_\parallel}{A_\perp}$. 

Imperfections in the HOM setup as well as the finite value of $g^{(2)}(0)$ influence the measured $V_\mathrm{raw}$. These imperfections can be accounted for in order to determine the true overlap $V$ of two single photon states produced by the source. If $P_2$ is the probability of creating two photons with one laser pulse, $P_{1}$ the probability of creating a single photon and $P_{0}$ the probability of creating the vacuum state, then $V$ can be calculated from $V_\mathrm{raw}$ under the assumptions that $P_{2} \ll P_{1} \ll P_{0}$ and that the two photons in the two-photon pulse are distinguishable \cite{Santori2002}. In principle, further corrections arise in the case $P_{2} \ll P_{1}$ but $P_{1} \ge P_{0}$, as achieved at the output fibre of the experiment. (An additional HOM signal arises when a two-photon and a single-photon pulse are created successively.) In practice however, the HOM setup has a low throughput and hence the assumption $P_{2} \ll P_{1} \ll P_{0}$ is reasonably fulfilled in the HOM measurements.The result is:
\begin{equation}
\label{eq:correction_formulat}
\begin{aligned}
&
V=\frac{1}{\left(1-\epsilon\right)^2}\left(1+2g^{(2)}\left(0\right)\right)\left(\frac{R^2+T^2}{2RT}\right)V_\mathrm{raw},
\end{aligned}
\end{equation}
where $T$ and $R$ are the transmission and reflection coefficients of the fibre beam-splitter, and $(1-\epsilon)$ is the classical visibility of the interferometer. Assuming further that $R$ and $T$ are close to 50\%,
\begin{equation}
\label{eq:correction_formulat_2}
V=\frac{1}{\left(1-\epsilon\right)^2}\!\left(1+2g^{(2)}\left(0\right)\right)\!\left(1+2\left(R-T\right)^2\right)V_\mathrm{raw}.
\end{equation}
We characterised the optical setup and extracted $R = 0.495$, $T = 0.505$ and $(1-\epsilon)=0.995\pm0.0025$. The correction due to the imbalance in the beam-splitter is negligible as the splitting ratio is close to 0.5:0.5 such that the main contributions to the correction arise from the limited visibility of the interferometer and the small but finite $g^{(2)}(0)$ of the source.

Supplementary Fig.~\ref{fig:HOM}b shows the raw and the corrected HOM visibilities as a function of the delay in the interferometer. The error bars on $V_\mathrm{raw}$ represent the errors arising from the shot noise of the detector. The error bars on $V$ include in addition uncertainties in the exact visibility of the interferometer and uncertainties in $g^{(2)}(0)$. Within error, $V$ is insensitive to the time delay between the interfering photons, indicating that the coherence time of the source is significantly longer than 1.5~$\mu$s.

\section{Numerical simulation of the microcavity}
\subsection{Calculation of the $\mathcal{Q}$-factor}

\begin{figure*}[t!]
  \centering
  \includegraphics[width=120 mm]{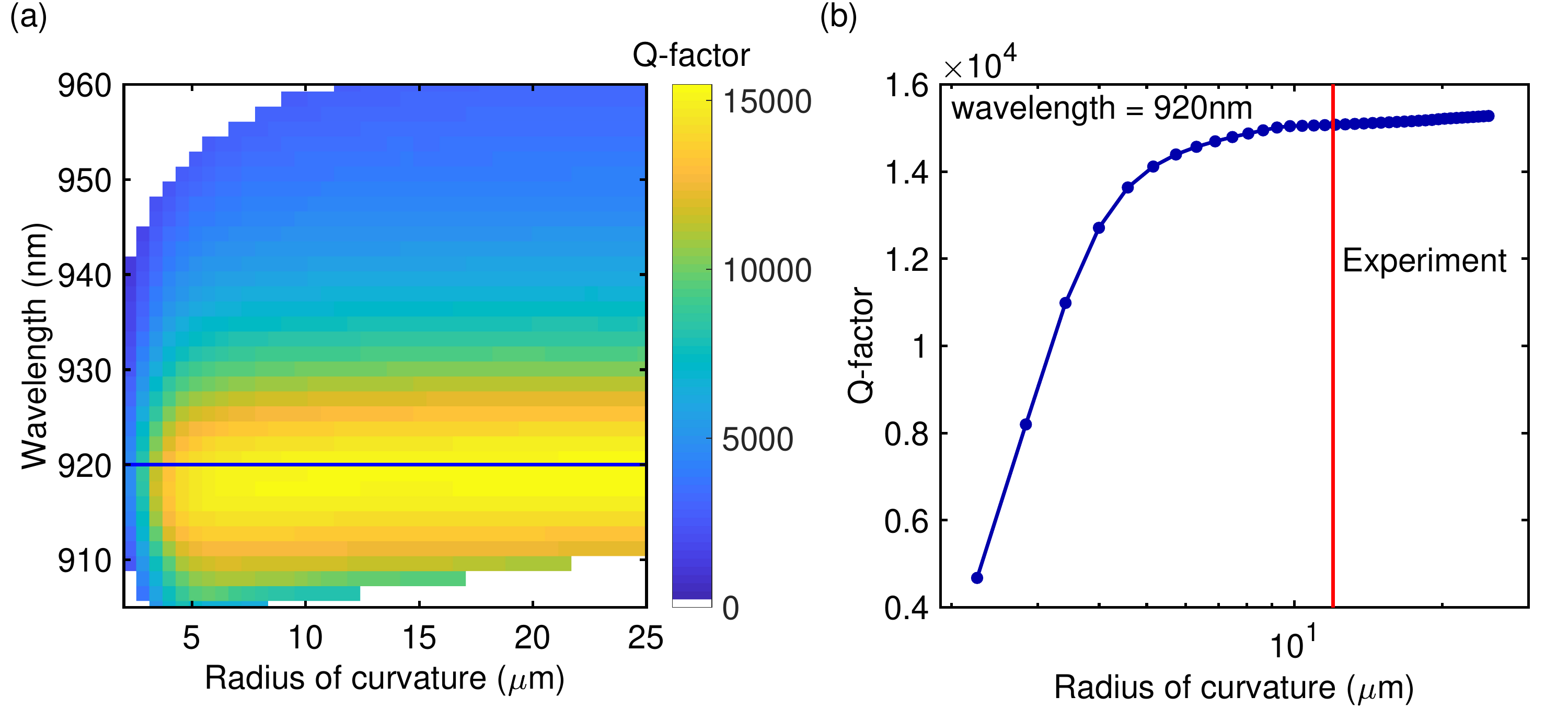}\\
  \caption{\footnotesize
  \textbf{Effect of diffraction losses on the $\mathcal{Q}$-factor.} (a) Calculated $\mathcal{Q}$-factor as a function of the wavelength and the radius of the curvature of the top mirror. The $\mathcal{Q}$-factor is maximum at the centre of the stopband (920\,nm). (b) A cut-through of the data at the wavelength of 920\,nm corresponding to the blue line in part (a). The $\mathcal{Q}$-factor drops significantly for radii smaller than 6\,$\mu$m signalling diffraction losses. For larger radii the $\mathcal{Q}$-factor approaches 15,200.}\label{fig:QvsR}
\end{figure*}

The microcavity $\mathcal{Q}$-factor was calculated using a one-dimensional transfer matrix simulation (The Essential Macleod, Thin Film Center Inc.). The top mirror is described using the design parameters taking the manufacturer's values for the refractive index (mirror design: silica-(HL)$^{7}$H with H (L) a quarter-wave layer in the high- (low-) index material at wavelength 920~nm, refractive indices 2.09 (1.48)). The transmission loss per round trip of the top mirror is 10,300 ppm. The bottom mirror has a nominal design GaAs-(HL)$^{46}$-active layer with H (L) a quarter-wave layer in GaAs (AlAs) at wavelength 940~nm, as shown in Supplementary Fig.~\ref{fig:mode}a. In practice, the layers become gradually thinner during growth. The wavelength of the stopband and the oscillations in reflectivity out with the stopband can be very well described by postulating a linear change in thickness during growth \cite{Najer2019}. The losses in the entire semiconductor heterostructure (including the free-carrier absorption in active layer) can be assessed by measuring the $\mathcal{Q}$-factors with an extremely reflective, extremely low-loss top mirror: the transmission loss is just 1 ppm per round trip; the absorption/scattering losses amount to 373 ppm per round-trip \cite{Najer2019}. These losses are negligible compared to the transmission loss of the top mirror. The simulated $\mathcal{Q}$-factor for the semiconductor DBR -- GaAs active layer (6 QWLs) -- air-gap (4 QWLs) -- top mirror structure is about 14,000. This is very close to the measured value, 12,600, taking here an average over the positions of the 6 QDs evaluated in this work, and averaging over the two microcavity modes.

Another possible source of losses in a microcavity is diffraction losses at the DBR mirrors, also termed ``side-losses" in the micropillar community \cite{Lalanne2004,Gregersen2010}.
For tightly confined modes, the angular spread in k-space expands, increasing the losses in the DBR mirrors and reducing the $\mathcal{Q}$-factor. We carried out numerical simulations to probe the effect of the radius of the curvature on the $\mathcal{Q}$-factor. Supplementary Fig.~\ref{fig:QvsR}a shows the $\mathcal{Q}$-factor as a function of the wavelength and $R$. Supplementary Fig.~\ref{fig:QvsR}b shows a cut-through of the data at the centre of the stopband (920\,nm). As expected, for small radii the  $\mathcal{Q}$-factor is a strong function of $R$ and drops to 4,600 at $R = 2.3\,\mu$m. At large radii ($R>6\,\mu$m), the $\mathcal{Q}$-factor is a weak function of R, and saturates at a value of 15,200. We use a top mirror with $R=11.98\,\mu$m in our experiments. We calculate a $\mathcal{Q}$-factor of 15,000 for this $R$ at the centre of the stopband, very close to the value at large radii. Hence, we conclude that side losses are negligible in our experiment. The $\mathcal{Q}$-factor calculated with the exact top mirror used in the experiment reduces to 14,000 due to slightly reduced reflectivity achieved during coating relative to the designed structure.

\subsection{Calculation of the QD-microcavity coupling}
In order to estimate the QD-microcavity coupling, a finite-elements method (Wave-Optics Module of COMSOL Multiphysics) is used to compute the vacuum electric field amplitude $|E_{\rm vac}(r,z)|$ confined by the microcavity (Supplementary Fig.~\ref{fig:mode}b). The model assumes axial symmetry about the optical axis ($(x,y)=0$). We use a 1\,$\mu$m thick perfectly index-matched layer at all outer boundaries of the simulation to prevent internal reflections. The model takes a top mirror with radius of curvature $R = $ 11.98\,$\mu$m and sagittal height $s = $ 0.41\,$\mu$m, exactly the mirror used in the experiments (see Sec.~\ref{sec:craterfab}). This model also allows the $\mathcal{Q}$-factor to be determined, yielding $\mathcal{Q}=14,000$, in agreement with the one-dimensional transfer matrix calculation. $\mathcal{Q}=14,000$ corresponds to $\kappa/(2\pi) = 23.3$\,GHz where $\kappa$ is the decay rate of the microcavity. At the location of the QDs ($z=z_{\rm QD}$) in the exact anti-node of the microcavity mode ($r=0$), the field is $|E_{\rm vac}(0,z_{\rm QD})|=35,000$ V/m. A QD at these wavelengths (920\,nm) has an optical dipole of $\mu/e = 0.71$\,nm where $e$ is the elementary charge \cite{Dalgarno2008}. The X$^{+}$ consists of two degenerate circularly-polarised dipole transitions (at zero magnetic field). We consider the interaction of one of these circularly-polarised dipoles with a linearly-polarised microcavity mode. The predicted QD-cavity coupling is therefore $\hbar g=\mu \cdot E_{\rm vac}(0,z_{\rm QD})/\sqrt{2}$ giving $g/(2\pi)=4.24$\,GHz. This dipole moment implies a natural radiative decay rate of 1.72\,ns$^{-1}$, equivalently $\gamma/(2\pi)=0.27$\,GHz (assuming the dipole approximation in an unstructured medium). The calculated Purcell factor is therefore $F_{\rm P}=4g^2/(\kappa\gamma)=11.4$.

The Purcell factor and coupling $g$ can be determined from the experiment. We focus on QD1. The natural radiative decay $\gamma$ rate can be determined by gradually tuning the microcavity out of resonance with the QD, extrapolating the decay rate to large detunings (main text, Fig.~2b). This gives $\gamma/(2\pi)=0.30$\,GHz. This agrees well with the estimate above. On resonance, the total decay rate increases to 3.33\,GHz. In the experiment however, the polarisation-degeneracy of the microcavity is lifted (see Sec.~\ref{characterisation}) and the QD exciton, an X$^{+}$, interacts with both microcavity modes. We focus on the resonance with the H-polarised mode. We determine the contribution to the total decay rate from the presence of the V-polarised microcavity mode by fitting the total decay rate as a function of microcavity detuning to two Lorentzians (main text, Fig.~2b). We subtract the contribution from the V-polarised mode and the free-space modes at the resonance with the H-polarised mode, giving a decay rate of $\gamma_{\rm H}/(2\pi)=2.87$\,GHz. This is the decay rate contribution we would expect if the V-polarised mode were highly detuned, in other words if the microcavity mode-splitting were very large. At this limit, a circularly-polarised dipole interacting with a single linearly-polarised microcavity mode, allows a comparison to be made with the calculated properties of the microcavity. The Purcell factor arising from the H-polarised mode alone is therefore $F_{\rm P}^{\rm H}=\gamma_{\rm H}/\gamma=9.6$, close to the calculated value (11.4). Using $F_{\rm P}^{\rm H}=4g^2/(\gamma \kappa)$ and taking $\kappa/(2\pi)=24.00$\,GHz, the experimental value for the H-polarised mode (wavelength 919 nm, $\mathcal{Q}=13,600$), we determine $g/(2\pi)=4.16$\,GHz. This is close to the calculated value (4.24\,GHz). (Exact agreement is not expected as the QD dipole fluctuates from QD to QD.) We can conclude that, first, the vacuum field in the real microcavity is compatible with the value calculated from the microcavity's geometry; and second, that the lateral tuning of the microcavity enables the QD to be positioned at the anti-node of the vacuum field.

\subsection{Properties of the output mode}
\label{subsec:na}
A simulation of the microcavity mode was used to determine the parameters of the output beam of the microcavity, notably the beam waist. The calculated beam in the SiO$_2$ substrate, i.e.\ in the region above the top mirror (Supplementary Fig.~\ref{fig:mode}c), is fitted to a Gaussian beam~\cite{Nagourney2010} of the form 
\begin{equation}
|E(r,z)|=|E_0|\frac{w_0}{w(z)}e^{-r^2/w^2(z)}
\end{equation}
with waist radius at $z$ given by
\begin{equation}
w^2(z)=w_0^2\left(1+\left(\frac{z}{z_{\rm R}}\right)^2\right).
\end{equation}
$z_{\rm R}=n\pi w_0^2/\lambda_0$ is the Rayleigh range in the medium (refractive index $n=1.4761$ is taken for SiO$_2$). The fit taking $w_0$ (and $|E_0|$) as fit parameters results in $w_0=1.05$\,$\mu$m. This corresponds to a numerical aperture of ${\rm NA}=\lambda_0/(\pi w_0)=0.279$.

\section{Optical setup}
In the experiment, the microcavity and one lens, the objective lens, are mounted in a helium bath-cryostat ($T=4.2$\,K). A window enables free optical-beams to propagate from an optical setup at room temperature to the microcavity system at low temperature \cite{Barbour2011,Greuter2014,Greuter2015,Najer2019}, as shown in Supplementary Fig.~\ref{fig:setup}. The top-mirror of the microcavity is fixed at the top of a titanium ``cage", inside which the sample, mounted on a piezo-driven XYZ nano-positioner, is placed \cite{Barbour2011,Greuter2014,Greuter2015,Najer2019}. The nano-positioner allows for full \textit{in situ} spatial (XY) and spectral (Z) tuning of the microcavity. The titanium cage sits on another XYZ nano-positioner, which allows for positioning of the microcavity relative to the objective lens, an aspheric lens of focal length $f_{\rm obj}=4.51$\,mm (355230-B, ${\rm NA}=0.55$, Thorlabs Inc.), leading to close-to-perfect mode matching of the microcavity and the microscope. The microscope has a polarisation-based dark-field capability \cite{KuhlmannRSI2013}. As shown in Supplementary Fig.~\ref{fig:setup}, laser light is input into the microscope via a single-mode fibre. The beam is collimated by a $f_{\rm fibre}=11$\,mm aspheric lens (60FC-4-A11-02, Sch\"{a}fter + Kirchhoff GmbH). A linear polariser (LP) guarantees the polarisation-matching of the input beam to a polarising beam-splitter (PBS) which reflects the light towards the microcavity. A half-wave plate allows the axis of the polarisation to be rotated: the output state is chosen to match one of the principal axes of the microcavity, the V-axis. The light is then coupled into the microcavity by the objective lens. The same lens collects the microcavity output. H-polarised light is transmitted by the PBS and focussed by a lens (60FC-4-A11-02, Sch\"{a}fter + Kirchhoff GmbH) into a single-mode optical fibre (780HP fiber, Thorlabs Inc). In the dark-field scheme, the suppression of V-polarised laser light is optimised by adjusting an additional quarter-wave plate in the main beam-path. Confocal detection is crucial. For pulsed excitation, an extinction ratio up to $10^6$ is achieved (for continuous wave excitation, it is up to $10^8$) and remains stable over many days of measurement \cite{KuhlmannRSI2013}. 

The estimation of the microcavity beam waist (Sec.~\ref{subsec:na}) was used to optimise the fibre-coupling efficiency by selecting an appropriate aspheric lens in front of the optical fibre. The objective lens (355230-B, ${\rm NA}=0.55$, Thorlabs Inc.) has a focal length $f_{\rm obj}=4.51$\,mm. Its NA is considerably larger than the NA of the microcavity in order to minimise clipping losses. The lens coupling the output into the final optical fibre should be chosen to ensure mode-matching with the single-mode in the fibre. The fibre has a nominal mode-field radius of $w_1=(2.71\pm0.27)$\,$\mu$m at $\lambda_0=920$\,nm (780HP fiber, Thorlabs Inc.). The focal length for optimum fibre-coupling is $f_{\rm fibre}=f_{\rm obj} \cdot w_1/w_0=(11.6\pm1.2)$\,mm. Thus, an $f_{\rm fibre}=11$\,mm aspheric lens was chosen for the experiments.

The mode-locked laser (Mira 900-D picosecond mode, Coherent GmbH) operates at a repetition rate of 76.3\,MHz. The spectral width lies in the range between 60 and 100\,GHz corresponding in the transform-limited case to temporal widths between 5 and 3\,ps, respectively. The temporal width is the full-width-at-half-maximum of the intensity.

\begin{figure}[t!]
	\includegraphics[width=80 mm]{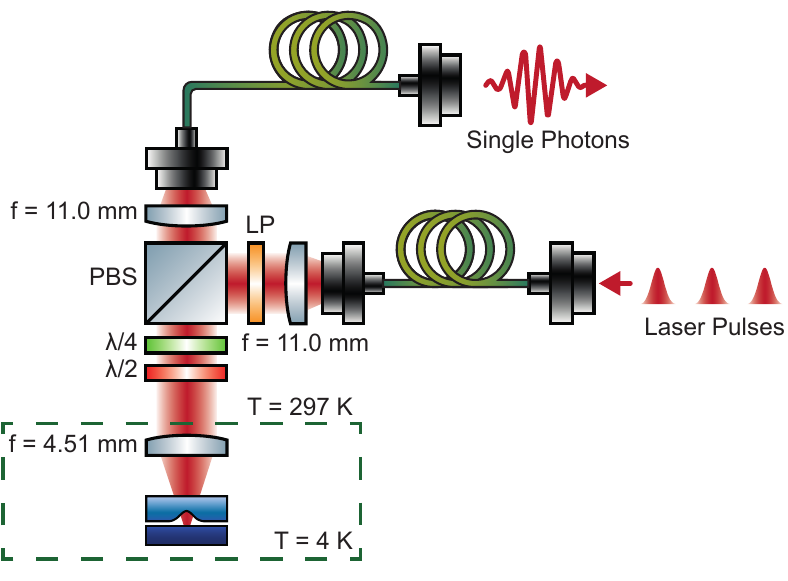}
	\begin{singlespace}
 \caption{\footnotesize
 \textbf{The optical setup.}
 \textnormal{The microcavity system resides in a cryostat at $T=4.2$\,K. Light is coupled in and out of the microcavity with a polarisation-based dark-field microscope. The objective lens is placed inside the cryostat along with the microcavity; the rest of the microscope is located outside the cryostat. Laser light enters via a single-mode optical fibre and is collimated with an $f=11$ mm lens, passing through a linear-polariser (LP). The input is reflected by a polarising beam-splitter (PBS); the polarisation axis of the excitation, the V-axis, is set by the half-wave plate ($\lambda/2$). The PBS and a quarter-wave plate ($\lambda/4$) suppress the coupling of unwanted back-reflected laser light into the collection arm. H-polarised single photons generated by the emitter are transmitted through the PBS and focused into the final single-mode optical fibre.}
 }
 \label{fig:setup}
 \end{singlespace}
\end{figure}

\section{Calibration of Detectors}
Two photon-counting detectors were used to perform experiments in this work, a superconducting NbTiN-nanowire single-photon detector (SNSPD) unit (EOS 210 CS Closed-cycle, Single Quantum B.V.) optimised for operation at 950 nm; and a near-infrared optimised, fibre-coupled silicon avalanche photodiode (APD, model SPCM-NIR, Excelitas Technologies GmbH \& Co.\ KG). In order to determine the efficiency of single-photon creation in this work, a careful calibration of the detectors' efficiencies was performed.

The measurement relies on a setup with a free-space laser beam (out-coupled from an optical fibre with angled facet), a set of calibrated neutral density filters (NDs) that can be placed in and out of the beam path, and a second optical fibre into which the beam is coupled (in-coupling via an angled facet). The frequency $\nu$ of the laser light is determined precisely prior to measurement with a interferometric device (HighFinesse Laser and Electronic Systems GmbH). For optical power $P$, the photon flux is $\frac{P}{h \nu}$ where $h$ is Planck's constant.

With the NDs removed from the beam's path, the optical power emerging out of the second fibre is measured with a calibrated silicon photodiode (Sensor Model S130C, Power measuring console PM100D, Thorlabs Inc.). The attenuating NDs are subsequently placed into the beam's path in order to avoid saturating the photon-counting detectors. The photon rate out of the fibre is then measured using both the SNSPD and the APD. The efficiency of each detector is given by the ratio of the measured count-rate to the known photon flux.

The efficiency of the SNSPD is determined to be $\eta_{\rm SNSPD} = $(82$\pm$5)\%. This value matches closely the specifications provided by the manufacturer at a wavelength of 940\,nm. The efficiency of the APD is $\eta_{\rm APD} = $(42$\pm$3)\% with an angled facet directly in front of the detector (FC-APC type fibre). The efficiency is slightly higher, $\eta_{\rm APD} = $(44$\pm$3)\%, with a flat facet directly in front of the detector (FC-PC type fibre). The errors in the measurements arise from 4\% in the calibration of the NDs, 1.5\% in the calibration of the NDs, 3\% nominal error of the silicon photodiode, and shot noise in the detectors (1.0\%).

We note that for the APD, due to the dead-time of the detector (typically $\sim 20$\,ns), a linearity correction factor must be applied to count rates above 200\,kHz. This correction factor scales quadratically from 1 at 200\,kHz to 3.32 at 25\,MHz. For the experiments performed in this work, the appropriate correction factor was applied to take this effect into account. It results in a change in efficiency of a few \% at the count rates in Fig.~2c (main paper).

\section{Theory: Filtered Pulse Excitation Mechanism}
\label{sec:theory}

\begin{figure*}[t!]
        \includegraphics[width=120 mm]{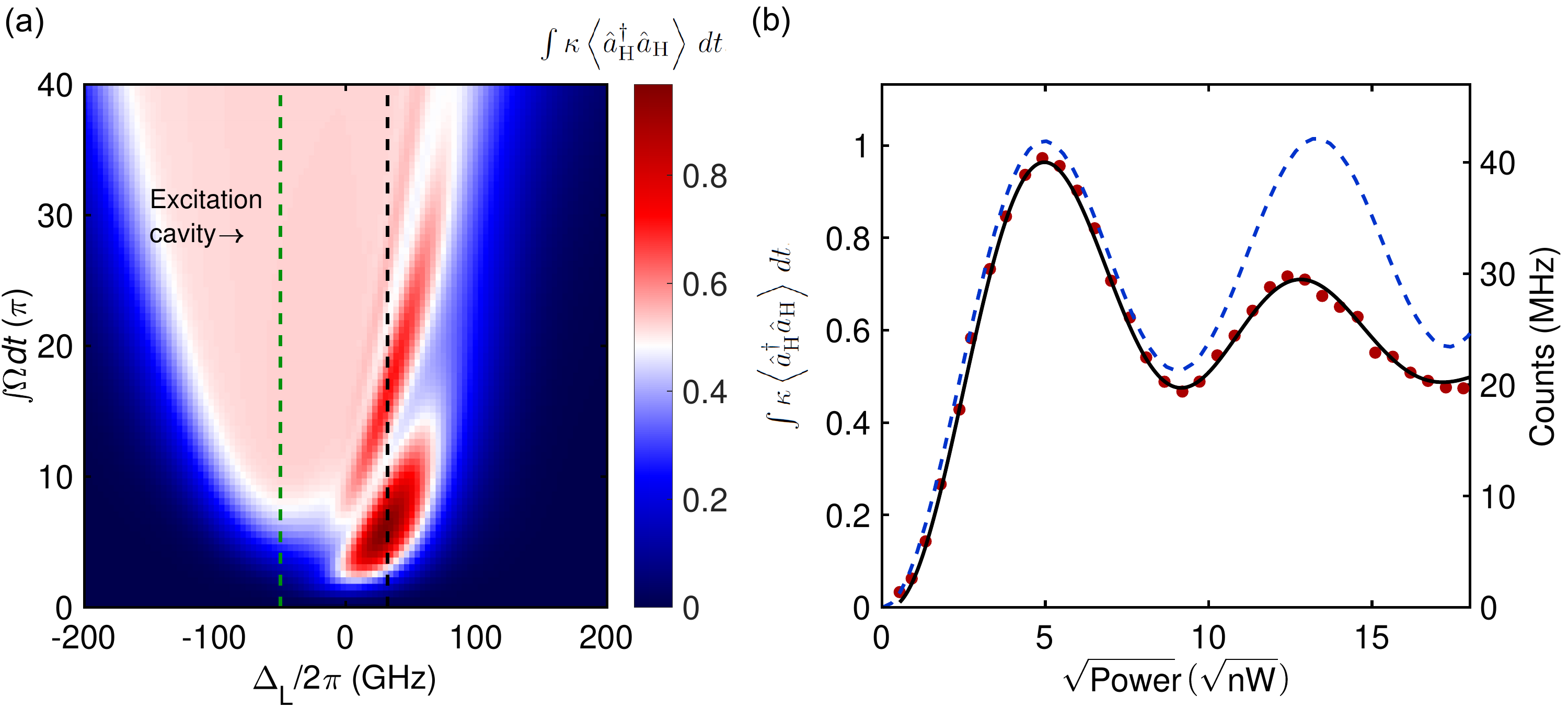}
	\begin{singlespace}
 	\caption{\footnotesize
 	\textbf{Calculated photon emission probability of a two-level system driven by filtered optical pulses.}
 	\textbf{(a)} \textnormal{The photon emission probability $\int\kappa\left\langle \hat{a}^\dag_{\rm H} \hat{a}_{\rm H}\right\rangle\, dt$ as a function of the laser detuning, $\Delta_{L}$, and the excitation power. For this simulation, $\kappa/(2\pi) = 25$\,GHz and $A = $32\,fs/K. The detuning between the excitation cavity and the TLS is 50\,GHz, as indicated by the green dashed line on the colour plot. \textbf{(b)} Photon emission probability as a function of power: the theory (black line) along with scaled experimental results (red points). The theoretical curve corresponds to the dashed black line in part \textbf{(a)}. The dashed blue line is the theory calculated with the same parameters except $A = 0$.}
	}
	\label{fig:cav_RF}
 \end{singlespace}
\end{figure*}

This section outlines the excitation mechanism and describes the theoretical model that is used to describe the experimental results. The Hamiltonian of a two-level system (TLS) interacting with a drive field and a resonant cavity mode, the H-polarised mode is given by:
\begin{multline}
\label{eq:TLS_H}
\hat{H} = \frac{\hbar\omega_0}{2}\hat{\sigma}_z+\frac{\hbar}{2} \left(\Omega^-(t)\hat{\sigma}_{+}+\Omega^+(t)\hat{\sigma}_{-}\right)+ \\
\hbar g\left(\hat{a}_{H}^{\dag}\hat{\sigma}_{-}+\hat{a}_{H}\hat{\sigma}_{+}\right),
\end{multline}
where $\hbar \omega_{0}$ is the energy difference between the excited state and the ground state of the TLS, $g$ is the coupling constant between the cavity and the TLS, $\Omega^\pm(t)$ are the positive and negative frequency components of the driving field, and $\hat{a}_{H}$ is the annihilation operator for the H-polarised cavity mode. 

In this work, the TLS is resonant with the H-polarised cavity mode but the optical pulses enter the cavity via the red-detuned V-polarised cavity mode. The optical pulse (frequency $\omega_{L}$) is blue-detuned (by frequency $\Delta_{L}$) with respect to the H-polarised cavity mode, $\Delta_{L} = \omega_L - \omega_0$. (The scheme is shown in Fig.~1c of the main paper.) As such, the TLS is driven not by the bare optical pulse but by the pulse modified by the off-resonant V-polarised cavity mode. $\Omega^\pm(t)$ can be calculated by convoluting the optical pulse with the impulse response function of a cavity, i.e.\ $e^{(-\kappa t/2)}\mathrm{cos}(\omega_c t)$, retaining the positive (negative) frequency components. ($\omega_c$ is the angular-frequency of the cavity, here that of the V-polarised cavity mode.) We introduce leakage out of the cavity mode with the Lindblad operator $\hat{\mathscr{L}}=\sqrt{\kappa}\,\hat{a}_{\rm H}$, where $\kappa$ is the decay rate of the cavity mode. Phonon-induced dephasing is modeled with $\hat{\mathscr{L}}=\sqrt{AT/2}\,|\Omega(t)|\,\hat{\sigma}_{z}$, where $T$ is the temperature of the phonon bath, and $A$ is the parameter describing the interaction between the phonons and the exciton \cite{Ramsay2010PRL}. Finally, the photon emission probability is calculated as $\int\kappa\left\langle \hat{a}^\dag_{\rm H} \hat{a}_{\rm H}\right\rangle\, dt$.

We use the Python package Qutip \cite{JOHANSSON2012CMPT,JOHANSSON2013CMPT} to set up and solve the equations of motion based on the Hamiltonian in Eq.~\ref{eq:TLS_H}. Supplementary Fig.~\ref{fig:cav_RF}a shows the photon emission probability as a function of the laser frequency and excitation power. In these simulations, the excitation cavity is $-50$\,GHz away from the TLS resonance corresponding to the mode-splitting of our cavity (QD1), and the pulse width is $t_p = $\,5.2\,ps (defined as the full-width-at-half-maximum of the laser intensity). The photon emission probability shows Rabi-like oscillations as expected from a driven TLS. However, as evident from the plot, the full inversion of the TLS takes place when the excitation laser is blue-detuned from the TLS resonance contrary to the textbook case where maximum inversion is obtained under strict resonance conditions. The crucial point is that the incomplete Rabi oscillations do not signify an incomplete inversion of the TLS. On the contrary, by choosing the optimal detuning of the laser pulse, the photon emission probability reaches values close to one, Supplementary Fig.~\ref{fig:cav_RF}b.

We model the experimental results with the following parameter set: $\Delta_L/(2\pi) =  32$\,GHz, $t_p = 5.2$\,ps, and $A = 32$\,fs/K. These parameters match well with the expected pulse-width based on the measured spectral width and the detuning $\Delta_L/(2\pi) =  32$\,GHz found in the experiment. There is a mismatch between the phonon-induced dephasing rate and the value reported in Ref.~\cite{Ramsay2010PRL}. However, such a discrepancy is not uncommon in the literature \cite{Ramsay2010PRL,Ulrich2011,Monniello2013}, and could hint at other excitation-induced dephasing mechanisms. With these parameters, the model shows that a near complete inversion of the TLS is possible. As shown in Supplementary Fig.~\ref{fig:cav_RF}b, the calculated photon emission probability is 96.3\% at the first Rabi peak. Excitation-induced dephasing is important to describe the subsequent Rabi peak but results in only a small decrease at the first Rabi peak.

\bibliographystyle{naturemag_noURL}
\bibliography{SPS_bib_vArXiv}

\end{document}